\documentclass[namedreferences]{kluwer}
\usepackage[dvips]{epsfig}
\begin{document}
\begin{article}
\begin{opening}

\title{The uncertainties in the synthetic indices for stellar populations}
\author{Claudia \surname{Maraston}\email{maraston@usm.uni-muenchen.de}} 
\institute{Universit\"ats-Sternwarte M\"unchen}
\author{Laura \surname{Greggio}\email{greggio@usm.uni-muenchen.de}}
\institute{Osservatorio Astronomico Universit\`a di Bologna\\
{Universit\"ats-Sternwarte M\"unchen}}
\author{Daniel \surname{Thomas}\email{daniel@usm.uni-muenchen.de}}
\institute{Universit\"ats-Sternwarte M\"unchen}
\runningauthor{Claudia Maraston, Laura Greggio, and Daniel Thomas}
\runningtitle{The uncertainties in the synthetic indices for stellar populations}
\begin{abstract} 
The study of line strengths in the spectra of early-type galaxies has
proven to be a powerful tool for investigating the age and the
metallicity of these systems. When computing models for
spectrophotometric narrow-band indices, index calibrations as functions
of the relevant stellar parameters (e.g. temperature, gravity and metal
content) are used. Thus synthetic indices depend upon
these calibrations (called {\it fitting functions}), as well as on the
stellar evolution ingredients adopted. All these inputs suffer from
uncertainties, which impact on the derived value for the indices. In
this paper we address this problem quantitatively.

We compute synthetic ${\rm {Mg}_2}$, {\rm Fe5270}, {\rm Fe5335} and
${\rm H{\beta}}$ indices for Simple Stellar Populations (SSP) of various
ages and metallicities, under different prescriptions for the fitting
functions. This allows us to estimate the impact of the uncertainties
in the fitting functions. By comparing our
models to those of other authors computed with the same fitting functions,
we estimate the uncertainty associated to the use of different stellar
evolution prescriptions. It is found that the modelling of the Horizontal
Branch impacts particularly on {\rm Fe} and ${\rm H{\beta}}$. In the
range of parameters explored, the uncertainties introduced by
the use of different fitting functions can be appreciably larger than the
error affecting the observational data. This typically occurs at high $Z$
for the metallic line strengths, at low $Z$ for the ${\rm H{\beta}}$ index.
\end{abstract}
\end{opening}

\section{Introduction}

Defined either as equivalent widths or magnitudes, Spectral Indices
(SI) trace the strength of particular absorption features and are
widely used as age ($t$) and metallicity ($Z$) indicators for stellar
systems (e.g. elliptical galaxies, see \opencite{faber95}, 
\opencite{ffi95}, \opencite{b95} and \opencite{lau97}). In
general, a single SI depends on both parameters $t$ and $Z$ in the
same way: metallic lines become stronger with increasing age and
$Z$, while Balmer lines become weaker. As a consequence, a given value
for the SI corresponds to either relatively young ages and high
metallicities, or vice versa, which is known as the age/metallicity 
degeneracy (\opencite{w92}). However, since the H${\beta}$ line
strength is more sensitive to age, while metallic indices are more
sensitive to $Z$, these two fundamental parameters are derived by
analysing observational data in the 2-dimensional diagram {\rm
H${\beta}$} versus metallic line strengths (e.g. \opencite{gonza93}; 
Kuntschner, this volume). Errors in the data are usually taken into
account (see \opencite{trager98}) whereas uncertainties in the models
are not considered. 
The aim of this paper is to estimate these uncertainties.

We examine SI for SSPs (i.e. single age and single metallicity
populations of single stars), which are obtained by summing up the 
contributions from all the stars that compose the population. The 
$\rm Mg_{2}^{\rm SSP}$, for example, is given by 
\begin{equation} 
\rm Mg_{2}^{\rm SSP}(t,[{\rm Fe/H}])=-2.5\cdot\log{{\sum_{*} 
10^{-0.4\cdot({\rm Mg}_{2}^{*})}\cdot f{\rm_c^*}}\over{\sum_{*} 
f{\rm_c^*}}}
\end{equation}
In eq. 1, ${\rm f}_c^*$ is the continuum flux (in the relevant
wavelength window) and $\rm {Mg_{2}^{*}}$ is the value of the index
of a single star of the SSP. The latter is specified by
fitting functions, which depend on the stellar parameters
effective temperature, surface gravity and metal content
($T_{\rm eff}$, $g$, [Fe/H]), and are empirically calibrated on stellar
samples. Thus synthetic indices depend on: 1) the
fitting functions (FF); 2) the stellar evolution input (set of stellar
tracks, adopted law for the helium-enrichment parameter 
${\rm\Delta Y/\Delta Z}$, choice for the mass-loss rate, etc); 3) 
the Evolutionary Population Synthesis (EPS) computational procedure.

In the following we investigate the uncertainties by varying the above
input prescriptions.
  
\section{Results}

The SI pre\-sen\-ted here are com\-pu\-ted with the EPS
pro\-ce\-du\-re de\-scri\-bed in
\inlinecite{m98}, whi\-ch ma\-kes use of the Fuel Con\-sum\-ption Theo\-rem
(\opencite{rb86}) to eva\-lua\-te the ener\-ge\-tics of Post Main
Se\-quen\-ce (PMS) stars. The in\-put stel\-lar tra\-cks are taken from
\inlinecite{bono97} and from Cassisi (1998, {\it private communication}). In
order to estimate the impact of uncertainties in the FF, we explore the two
independent sets from Worthey et al. (1994, hereafter WFF) and from
Buzzoni et al. (1992; 1994, hereafter BFF). We selected the indices
common to the two sets, namely ${\rm Mg_2}$, ${\rm H{\beta}}$, {\rm
Fe5270} and {\rm Fe5335}. In addition, the comparison of our models
with those of Worthey (1994, hereafter WSSP) and of Buzzoni et al. (1992;1994,
hereafter BSSP) for the same FF, gives us clues on the uncertainties due to
the use of different EPS procedures and stellar evolution input.
We emphasize that the results obtained in this way provide lower
limits for the uncertainties affecting the model indices.

In this contribution we discuss part of the results obtained
in a comprehensive study that will be presented in a forthcoming paper 
(Greggio and Maraston, in preparation, hereafter GM99). 
We concentrate here on old ages, that are more relevant for
elliptical galaxies.

\subsection{The ${\rm Mg_2}$ index}

\begin{figure}
\centerline{
\epsfig{file=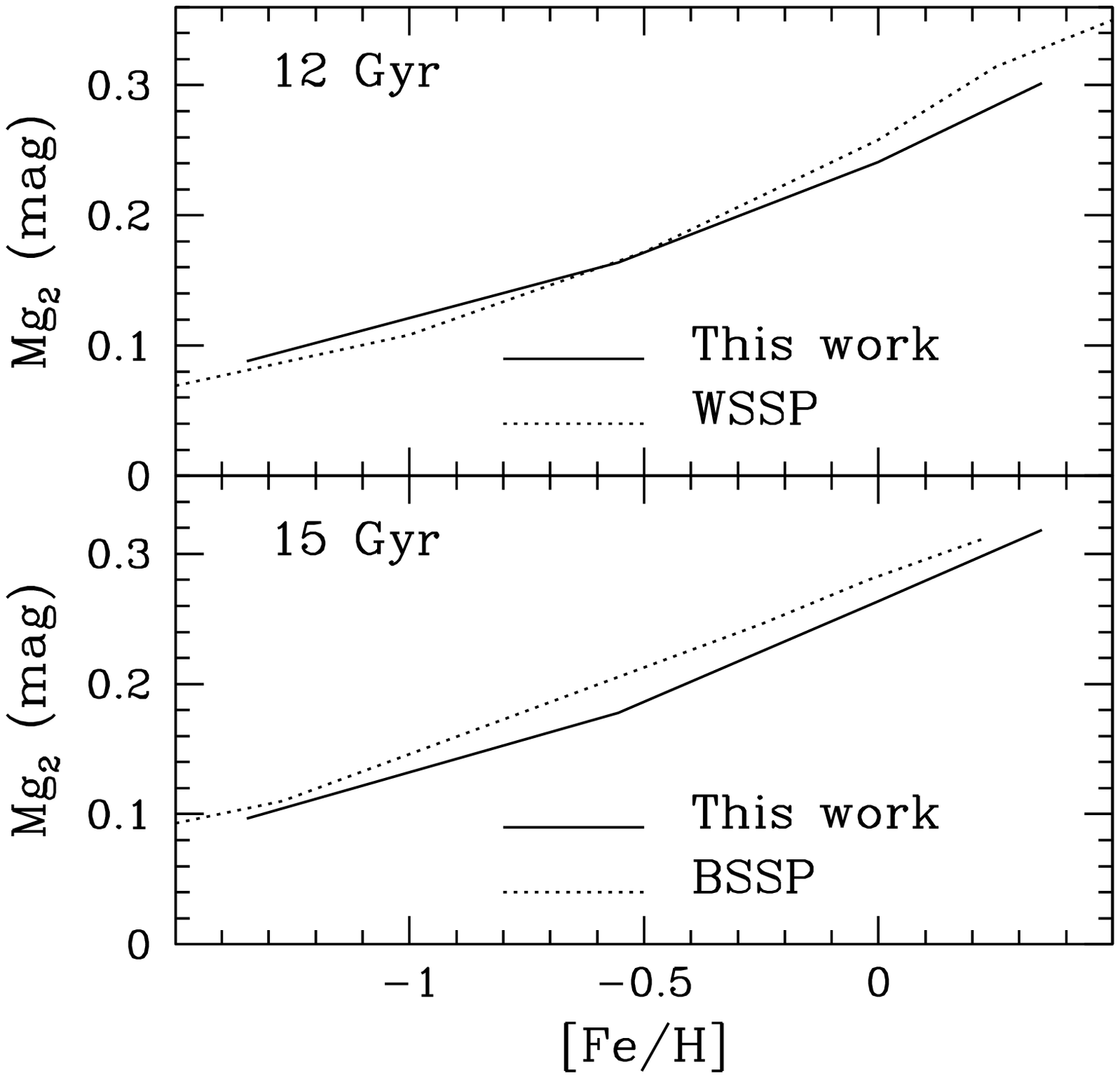,width=0.49\textwidth}
\epsfig{file=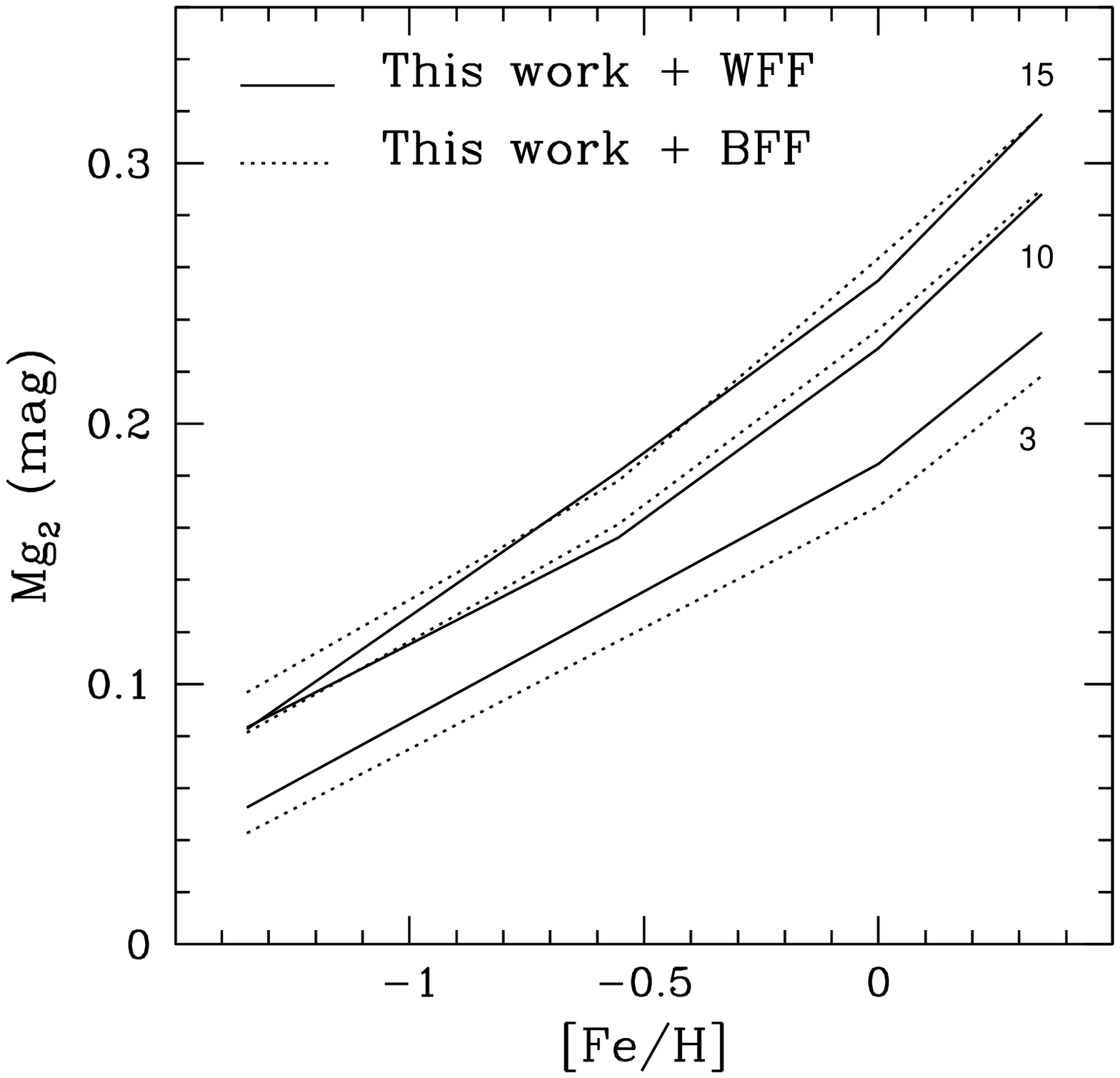,width=0.49\textwidth}}
\caption{{\em Left-hand panel}: the effect of different EPS. The upper
part of the diagram shows the comparison with WSSP at 12 {\rm Gyr}, the
lower part shows the comparison with BSSP at 15 {\rm Gyr}. {\em Right-hand
panel}: the uncertainty due to the FF. The curves are labelled with the SSP
ages in Gyr.}
\end{figure}
Figure 1 shows the effect of different EPS prescriptions (left-hand panel)
and of different FF (right-hand panel), on the synthetic ${\rm Mg_2}$. The
largest discrepancy due to the EPS is $\sim 0.02$ mag for
${\rm\-[Fe/H]}\sim\--0.5$ for the 15 Gyr old SSPs compared to Buzzoni's
models. When using the ${\rm\-Mg_2}$ index to estimate the metallicity {\rm
[Fe/H]} of a stellar system, the EPS procedure introduces an appreciable
uncertainty. In the cases explored here, the largest values are 0.2 dex, for
BSSP at ${\rm Mg_2}\sim 0.2$, and $\sim 0.15$ dex for WSSP at
${\rm\-Mg_2}\sim 0.3$. The ${\rm Mg_2}$ indices in BSSP models are
systematically stronger than ours. This is likely due to the use of a
different ${\rm\Delta Y/\Delta Z}$ of the evolutionary tracks used in the
two EPS, namely ${\rm\Delta Y/\Delta Z}\sim1$ in BSSP and
${\rm\Delta\-Y/\Delta Z}\sim 2.5$ in this work. At fixed stellar mass, a
lower $Y$ implies longer lifetimes on the Red Giant Branch (RGB) and shorter
lifetimes on the Horizontal Branch (HB) (see
\opencite{ren94}). Therefore, the indices of RGB stars receive relatively
more weight in eq. (1). Since the RGB is an important contributor to the
total ${\rm Mg_2}$ (see GM99), and since $\rm {Mg_{2}^{*}}$ becomes
stronger with decreasing $T_{\rm eff}$, an SSP with a lower
$Y$-content exhibits a stronger ${\rm\-Mg_2}$.

The discrepancies introduced by the different FF (Fig. 1, right-hand panel)
are relatively small at old ages (see also Table 1). For the 3 {\rm Gyr}
models, instead, WFF lead to ${\rm Mg_2}$ values which are systematically
higher by $\sim 0.015$ mag with respect to BFF. This comes from the higher
intrinsic ${\rm Mg_2}$ given by the WFF for the $T_{\rm eff}$ typical
of a 3 {\rm Gyr} turn-off ($\sim 6300~{\rm K}$). The largest uncertainty in
the {\rm [Fe/H]} determination hence appears for the 3 {\rm Gyr} models,
with {\rm $ \Delta {\rm [Fe/H]}\sim 0.15$}.

\subsection{The iron indices}

\begin{figure}
\centerline{
\epsfig{file=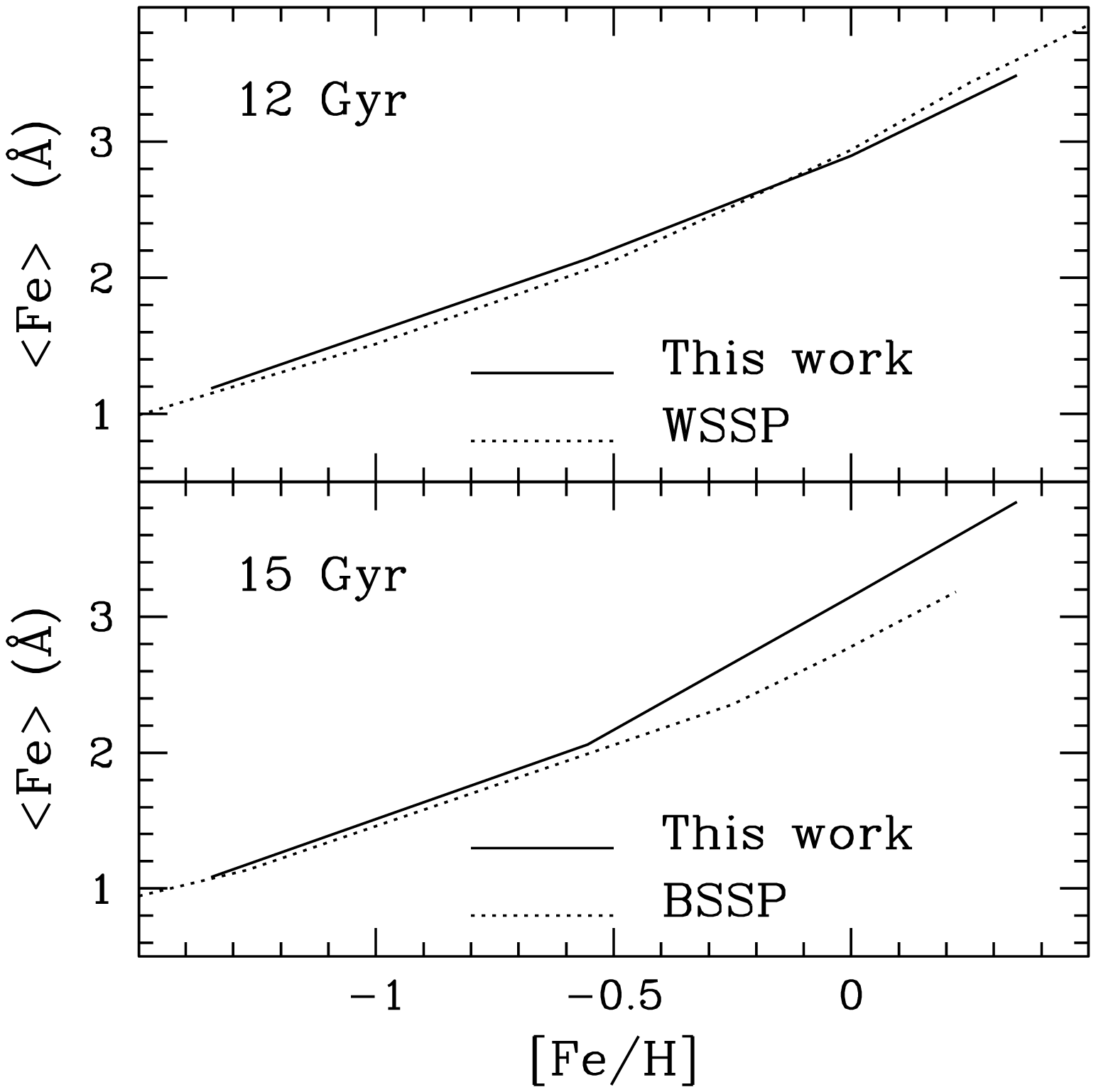,width=0.49\textwidth}
\epsfig{file=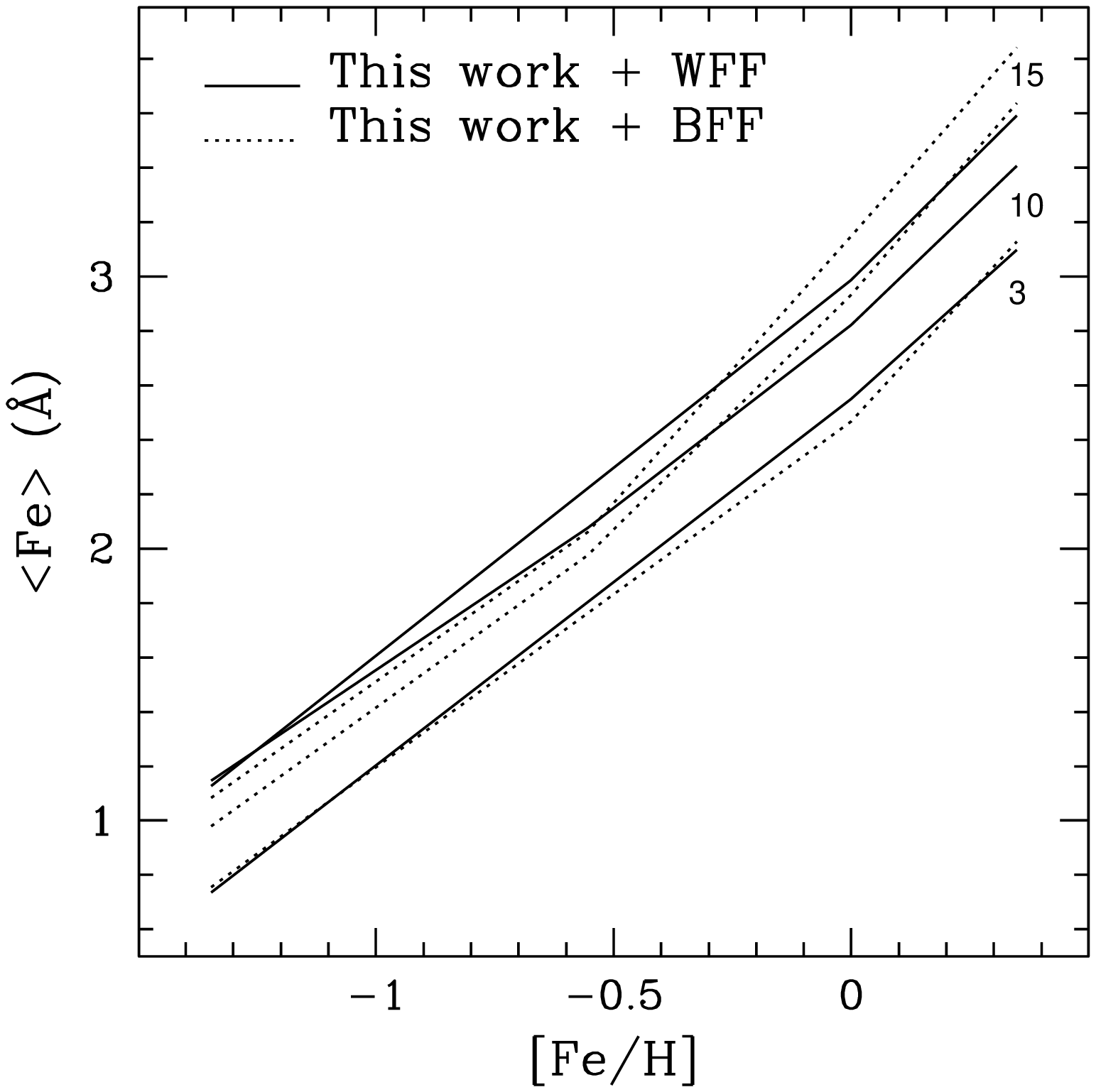,width=0.49\textwidth}}
\caption{{\em Left-hand panel}: the effect of different EPS. The upper
part of the diagram shows the comparison with WSSP at 12 {\rm Gyr}, the
lower part shows the comparison with BSSP at 15 {\rm Gyr}. 
{\em Right-hand panel}: the uncertainty due to the FF. The curves are
labelled with the SSP ages in Gyr.} 
\end{figure}
Figure 2 show\-s the sa\-me com\-pa\-ri\-sons as in Figure 1 for the
a\-ve\-ra\-ge i\-ron in\-dex $\langle {\rm Fe}\rangle$={\rm (Fe5335 +
Fe5270)/2}. Concerning the EPS procedure (left-hand panel), the discrepancy
to BSSP is systematically increasing with metallicity up to 0.4 {\rm \AA},
with our models predicting $\langle{\rm Fe}\rangle$ stronger than Buzzoni's.
Again, at least part of this inconsistency likely comes from the different
${\rm\Delta\-Y/\Delta Z}$ adopted. At ${\rm [Fe/H]}>-0.5$, the HB phase is
spent at $T_{\rm\-eff}\sim\-5300-4600~{\rm K}$, where the {\rm Fe} fitting
functions yield the strongest values for the indices \cite{b94}. It follows
that the {\rm Fe} indices are very sensitive to the lifetime of this phase.
As mentioned in the previous section, at fixed $Z$, a larger value of $Y$
implies a longer HB lifetime, which in turn determines stronger {\rm Fe}
indices. Thus for a given value of $\langle{\rm Fe}\rangle$ our models with
BFF yield a lower {\rm [Fe/H]}, compared to BSSP, with a
discrepancy reaching 0.2 dex at ${\rm [Fe/H]}>0$.

The different FF (right-hand panel) imply systematic differences in the
slopes of the relations index/metallicity at old ages. At large values of
$\langle{\rm Fe}\rangle$, BFF indices are stronger than those computed with
WFF (particularly for the Fe5270, see Table 1). The difference increases
with metallicity up to to $\Delta\langle{\rm Fe}\rangle\sim 0.2$ \AA\ and
mainly results from the different values of the FF. We finally notice that
within the explored range, the largest uncertainty in the metallicity
determination due to the use of the two sets of FF amounts to
$\Delta\-{\rm\-[Fe/H]}\sim\-0.15~{\rm dex}$ at $\langle{\rm Fe}\rangle\sim 3$.

\subsection{The H$\beta$ index}

\begin{figure}
\centerline{
\epsfig{file=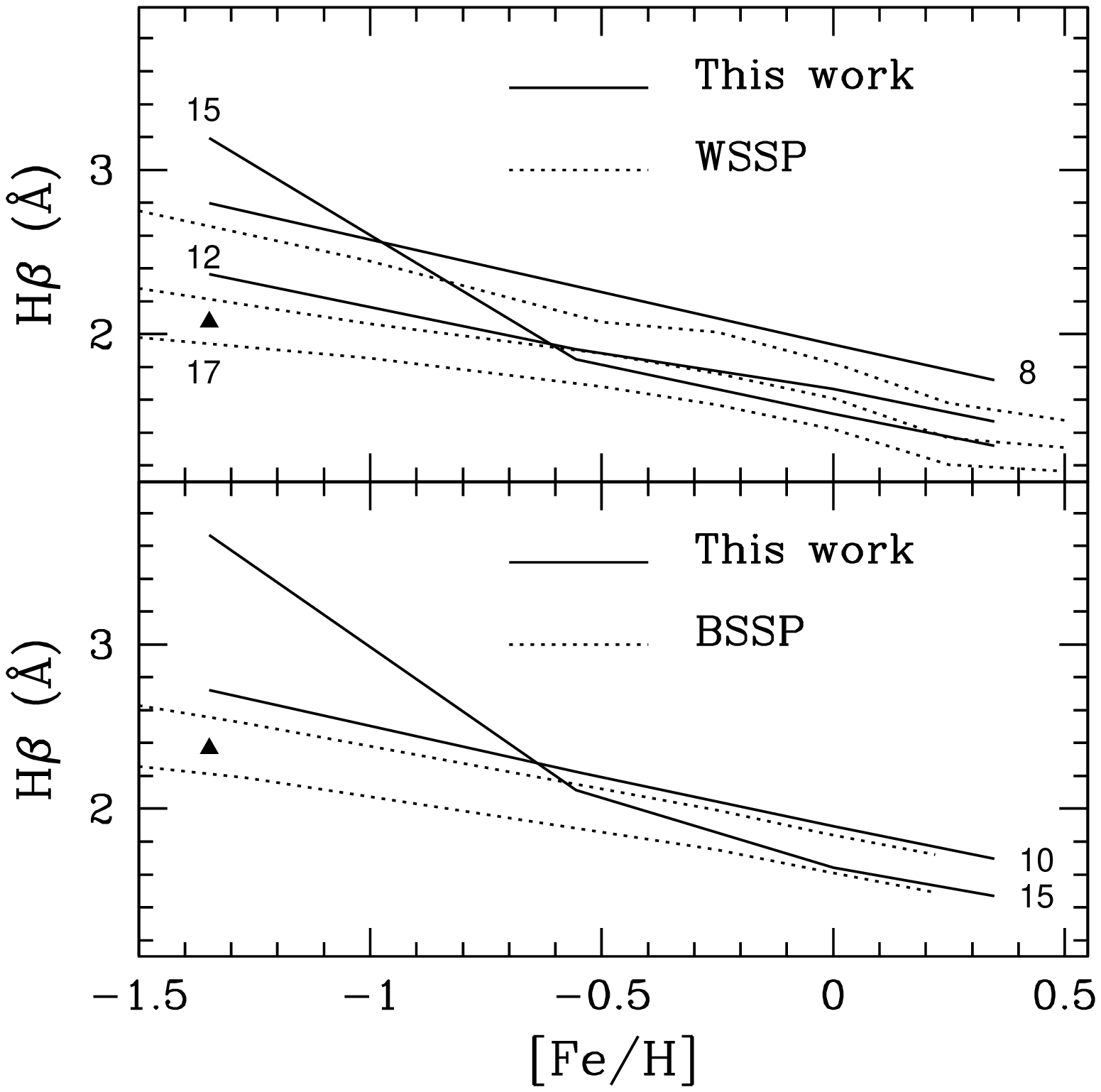,width=0.49\textwidth}
\epsfig{file=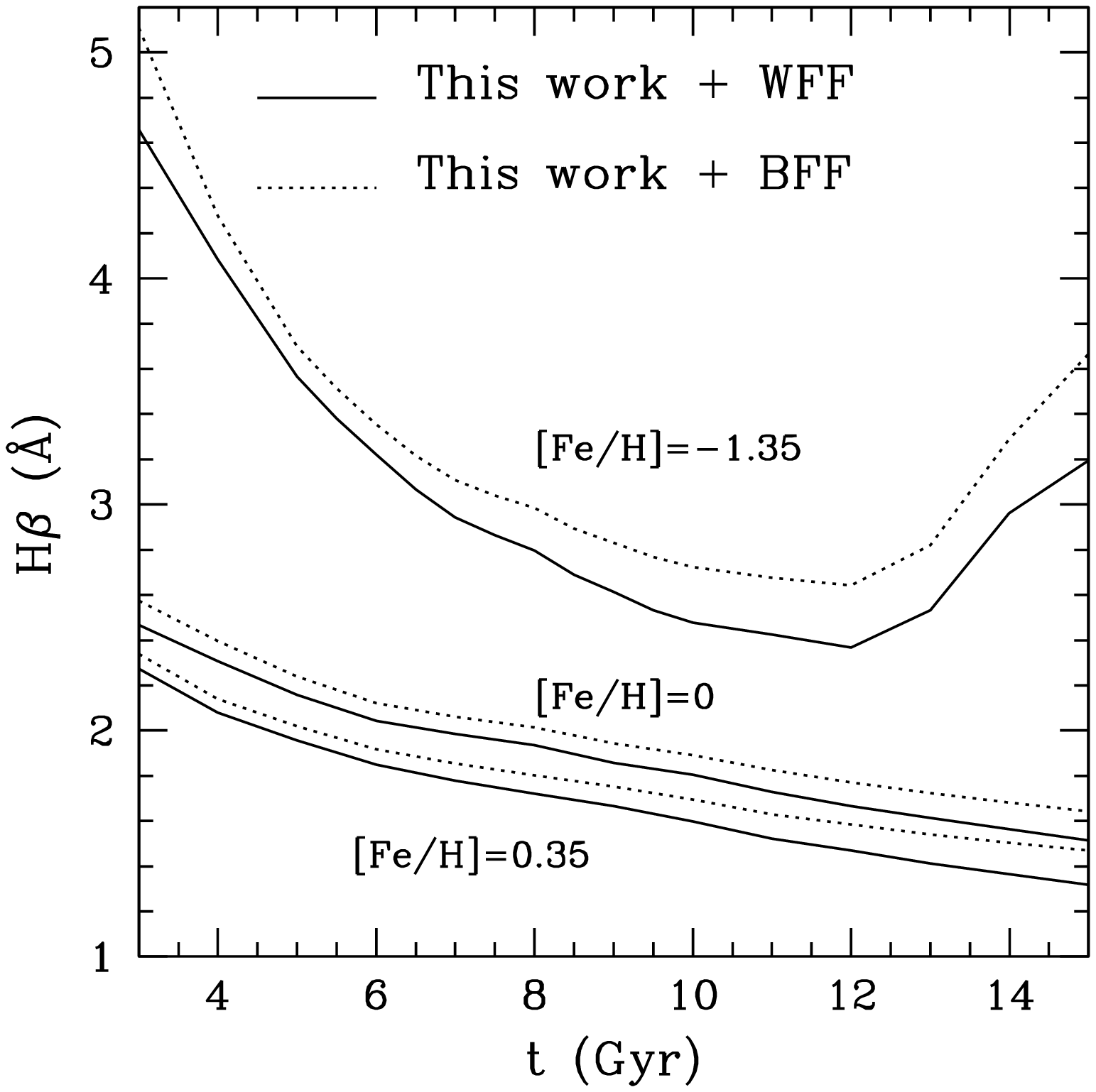,width=0.49\textwidth}}
\caption{{\em Left-hand panel}: the effect of different EPS.
In the top diagram we show our models for 8, 12 and 15 Gyr (solid lines),
and Worthey's models for 8, 12 and 17 Gyr (dotted lines). In the bottom
diagram ours and Buzzoni's models are shown for ages of 10 and 15 Gyr. In
both diagrams the triangles represent a model in which the mass-loss
during the RGB (see the text) is halved. {\em Right-hand panel}: H$\beta$
indices for two different choices for the fitting functions.}
\end{figure}
The comparison of the H$\beta$ indices from our models
with BSSP and WSSP is shown in the left-hand panel of Figure 3.
At low metallicities (${\rm [Fe/H]}~\lsim-0.5$ dex) and old
ages ($t~\gsim~12~{\rm Gyr}$), the H$\beta$ index is much
stronger in our computations. This large discrepancy comes from the
typical temperature of the HB phase in the various EPSs, which is
controlled by the assumptions on the mass-loss on the RGB. At a given
age (i.e. at fixed evolutionary mass at the turn-off), the larger
the mass-loss, the lower the mass of the HB star, which implies larger
$T_{\rm eff}$ for the HB phase.
In our models, the mass-loss is com\-pu\-ted u\-sing the sche\-me of 
\inlinecite{ml77} with the canonical efficiency $\eta =0.33$, a
value calibrated on Galactic globular clusters (see e.g. \opencite{fusi76}).
At 15 Gyr and ${\rm [Fe/H]}=-1.35$, the resulting evolutionary mass on the
HB is $\sim 0.65~M_{\odot}$ in our models, while Buzzoni
($\sim\-0.71~M_{\odot}$) and Worthey ($\sim 0.83~M_{\odot}$) use higher
values. We checked our computations comparing the $M_{\rm HB}$ values
with those obtained by Greggio and Renzini (1990) with the same $\eta$
values, and found no appreciable difference.

Thus, in our models the HBs are hotter, which leads to the strong
{\rm H${\beta}$} enhancement, since the relative FF are very sensitive to
$T_{\rm eff}$. Old and metal-poor (${\rm [Fe/H]}~\lsim -0.5$) SSPs
predominantly show this effect, while for higher metallicity, and younger
ages the evolutionary mass on the HB is relatively large, and the HB remains
at low temperatures ($T_{\rm eff}\sim 4600-4900$~K). However, the
$T_{\rm\-eff}$ of the HB depends on the prescriptions for the mass-loss, and
a larger value for $\eta$ would make the HB hotter also at higher $Z$ (see
e.g.
\opencite{gr90}).
Note that Worthey \& Ottaviani (1997) show an analogous effect of the RGB
mass-loss on the ${\rm H}\gamma$ and ${\rm H}\delta$ indices, as a
consequence of a change in the HB morphology with respect to Worthey
1994 (see the caption of their Fig. 6). Conversely, reducing the mass-loss on the RGB
leads to weaker {\rm H${\beta}$} indices at old ages/high metallicities, as
shown in Figure 3 by the model represented as a triangle. Therefore, {\rm
H${\beta}$} is not a reliable age indicator, its value also depending on the
HB morphology. Since the observed HB in Globular Clusters very often show
intermediate morphologies (see, e.g., \opencite{dickens72}), in GM99 the
results for mixed HB will be discussed.

The discrepancy in the synthetic {\rm H${\beta}$} due to the treatment of
the HB phase can be as large as $\sim$ 1.5 {\rm \AA} (see, e.g., the BSSP
case at ${\rm [Fe/H]}\sim -1.3, t = 15~{\rm Gyr}$). At low $Z$, this
ef\-fect im\-plies an age/me\-tal\-li\-ci\-ty de\-ge\-ne\-ra\-cy, with old
SSPs ($t\sim 12-14~{\rm Gyr}$) having the same {\rm H${\beta}$} index
of intermediate-age populations (see also \opencite{lau97}). This effect is
shown in the right-hand panel of Fig. 3, with the models at
${\rm [Fe/H]}\sim -1.35$ showing a minimum around 12 Gyr. Concerning the
uncertainties introduced by the FF, BFF systematically lead to higher index
values for all the explored values of age and {\rm [Fe/H]}. For
solar and supersolar metallicities, at ${\rm H{\beta}}\sim 1.5-1.6$,
models computed with WFF predict ages younger by $\sim 3~{\rm Gyr}$,
compared to models computed with BFF.

\section{Summary}

We have shown that current synthetic SI are affected by 
uncertainties that cannot be neglected when using these models to
derive the age and the metallicity of stellar systems. These
uncertainties originate from the EPS procedure/stellar
input and from the use of different Fitting Functions.

The $\rm Mg_{2}$ index seems to be the least affected by the different
modelling procedures. The EPS ingredient that mostly influence its values is
the ${\rm\Delta Y/\Delta Z}$ parameter, in the sense that at fixed $Z$ a
higher $Y$ leads to a lower $\rm\-Mg_{2}$. The dependence of the
$\rm\-Mg_{2}$ index on the adopted FF is found small at old ages, while for
intermediate age SSPs WFF predict systematically stronger values.

Also the $\langle{\rm Fe}\rangle$ index depends on ${\rm\Delta Y/\Delta Z}$, a
higher $Y$ yielding a stronger $\langle{\rm Fe}\rangle$. Models of the Fe
indices adopting different FF show major discrepancies at high {\rm
[Fe/H]} and old ages.

The {\rm H${\beta}$} index is highly sensitive to the treatment of
Horizontal Branch stars. In metal-poor SSPs (${\rm [Fe/H]}~\lsim -0.5$),
an age/metallicity degeneracy is found, since at $t~\gsim~12$ Gyr
{\rm H${\beta}$} becomes stronger with increasing age.
The sensitivity of this index on the FF is found to be large in all
the age and metallicity ranges explored, but especially in
the metal-poor regime.

Table 1 collects the es\-ti\-ma\-ted un\-cer\-tan\-ties due to dif\-fe\-rent
fit\-ting fun\-ctions, expressed as SI(WFF)$-$SI(BFF), for 3,10 and 15 {\rm
Gyr} old SSPs and (${\rm [Fe/H]}=-1.35,0.0,0.35$). The $2\sigma$
observational errors (from \opencite{gonza93}) are reported. Values in
bold-face are larger than the quoted observational errors. No value means a
negligible uncertainty ($\sim 10^{-3}$). There are regimes in which the
uncertainty related to the FF adopted in the modelling is appreciably
larger than the errors affecting the data. This happens at high $Z$ for
Fe5270, and at low $Z$ for the {\rm H${\beta}$} index. Thus, it appears that
uncertainties in the theoretical models should be taken into account when
deriving ages and metallicities for stellar systems.

To conclude, we stress that the estimates reported here are only indicative
of the real uncertainties affecting these indices. For example, a
subset of FF and SSP models existing in the literature is considered here.
Besides, variations in the parameters controlling the stellar evolution
input of EPS (e.g. $\eta$, $\Delta\-Y$/$\Delta Z$, IMF slope, etc.) have not
been discussed explicitly. A more thorough investigation will be the
subject of a forthcoming paper (GM99).

\begin{table}
\caption{Uncertainties on SI due to the use of different
fitting functions. Observational errors are from Gonz\'{a}lez 1993 (first line).
For the various ages and metallicities SI(WFF)$-$SI(BFF) is shown.}
\begin{tabular}{rrcccc}\hline
& & {\rm $\Delta {\rm Mg}_{2}$ (mag)} & {\rm $\Delta $ Fe5270 (\AA)} &
{\rm $\Delta $ Fe5335 (\AA)} & {\rm $\Delta$ H$_{\beta}$ (\AA)} \\
\multicolumn{2}{c}{Obs. $2\sigma$} & 0.014 & 0.12 & 0.14 & 0.12 \\
& & & & & \\
{[Fe/H]} & age & & & & \\
$-1.35$ & 3 Gyr & $+0.010$ & $-0.10$ & $+0.06$ & {\boldmath$-0.50$} \\
& 10 Gyr & - & $-0.12$ & {\boldmath $+0.22$} & {\boldmath $-0.25$} \\
& 15 Gyr & {\boldmath $-0.014$} & $-0.04$ & $+0.13$ & {\boldmath $-0.47$} \\
0.00 & 3 Gyr & {\boldmath $+0.016$} & $-0.02$ & {\boldmath $+0.18$} & $-0.11$ \\
& 10 Gyr & - & {\boldmath $-0.23$} & - & $-0.09$ \\
& 15 Gyr & - & {\boldmath $-0.27$} & $-0.05$ & {\boldmath $-0.13$} \\
0.35 & 3 Gyr & {\boldmath $+0.016$} & {\boldmath $-0.20$} & $+0.13$ & $-0.07$ \\
& 10 Gyr & - & {\boldmath $-0.39$} & $-0.08$ & $-0.10$ \\
& 15 Gyr &- & {\boldmath $-0.39$} & $-0.12$ & {\boldmath $-0.15$} \\ 
\hline
\end{tabular}
\end{table}

\begin{acknowledgements}
We greatly thank Santino Cassisi for kindly providing his carefully computed
stellar tracks and isochrones.
\end{acknowledgements}

\end{article}


\begin{thebibliography}{}


\bibitem[\protect\citeauthoryear{Bono et al.}{1997}]{bono97}
Bono, G., Caputo, F., Cassisi, S., Castellani, V. \& Marconi, M.
1997, ApJ, 489, 822


\bibitem[\protect\citeauthoryear{Buzzoni}{1995}]{b95}
Buzzoni, A., 1995, ApJS, 98, 69
 
\bibitem[\protect\citeauthoryear{Buzzoni et al.}{1992}]{b92}
Buzzoni, A., Gariboldi, G., and Mantegazza, L., 1992, AJ, 103, 1814

\bibitem[\protect\citeauthoryear{Buzzoni et al.}{1994}]{b94}
Buzzoni, A., Mantegazza, L., and Gariboldi, G., 1994, AJ, 107, 513

\bibitem[\protect\citeauthoryear{Chieffi and Straniero}{1989}]{oscar89} 
Chieffi, A. \& Straniero, O. 1989, ApJS, 71, 47

\bibitem[\protect\citeauthoryear{Dickens}{1972}]{dickens72} 
Dickens, R.,J., 1972, MNRAS, 157, 281 

\bibitem[\protect\citeauthoryear{Faber et al.}{1995}]{faber95} 
Faber, S. M., Trager, S.C., Gonz{\'a}lez, J.J., Worthey, G., in Van der
Kruit P. C., Gilmore G., eds, Proc. IAU Symp. 164, Stellar
Populations. Kluwer, Dordrecht, p. 249.  

\bibitem[\protect\citeauthoryear{Fisher, Franx \& Illingworth}{1995}]{ffi95} 
Fisher, D., Franx, M., Illingworth, G.,1995, ApJ, 448, 119

\bibitem[\protect\citeauthoryear{Fusi-Pecci \& Renzini}{1976}]{fusi76} 
Fusi-Pecci, F., Renzini, A., 1976, A\&A, 46, 447

\bibitem[\protect\citeauthoryear{Gonz{\'a}lez}{1993}]{gonza93} 
Gonz{\'a}lez J.J., 1993, PhD thesis, Univ. of California, Santa Cruz

\bibitem[\protect\citeauthoryear{Greggio}{1997}]{lau97}
Greggio, L., 1997, MNRAS, 301, 166

\bibitem[\protect\citeauthoryear{Greggio \& Renzini}{1990}]{gr90}
Greggio, L., Renzini, A., 1990, ApJ, 364, 35

\bibitem[\protect\citeauthoryear{Maraston}{1998}]{m98}
Maraston, C., 1998, MNRAS, 300, 872

\bibitem[\protect\citeauthoryear{Reimers}{1977}]{ml77}
Reimers, D., 1977, A\&A, 61, 217

\bibitem[\protect\citeauthoryear{Renzini}{1994}]{ren94}
Renzini, A., 1994, A\&A, 285, L5 

\bibitem[\protect\citeauthoryear{Renzini and Buzzoni}{1986}]{rb86}
Renzini, A. \& Buzzoni, A., in ``Spectral evolution of
Galaxies'', C. Chiosi and A. Renzini (eds.), Dordecht, Reidel, 1986, 195

\bibitem[\protect\citeauthoryear{Trager}{1998}]{trager98}
Trager, S.C., 1998, PhD thesis, Univ. of California, Santa Cruz

\bibitem[\protect\citeauthoryear{Worthey}{1992}]{w92}
Worthey, G., 1992, PhD thesis, Univ. of California, Santa Cruz

\bibitem[\protect\citeauthoryear{Worthey}{1994}]{w94m}
Worthey, G., 1994, ApJS, 95, 107

\bibitem[\protect\citeauthoryear{Worthey et al.}{1994}]{w94ff}
Worthey, G., Faber, S.M., Gonz{\'a}lez, J. and Burstein, D., 1994, ApJS,
94, 687

\bibitem[\protect\citeauthoryear{Worthey \& Ottaviani}{1997}]{wo97}
Worthey, G. and Ottaviani, D.,L., 1997, ApJS, 111, 377

\end{thebibliography}
\end{document}